
\def\J{$J/\psi$}

\def\P{$\psi'$}

\def\U{$\Upsilon$}

\def\c{c{\bar c}}
\def\b{b{\bar b}}

\def\t{\tau}

\def\Q{Q{\bar Q}}

\def\e{\epsilon}

\def\lsim{\raise0.3ex\hbox{$<$\kern-0.75em\raise-1.1ex\hbox{$\sim$}}}
\def\gsim{\raise0.3ex\hbox{$>$\kern-0.75em\raise-1.1ex\hbox{$\sim$}}}

\newcount\REFERENCENUMBER\REFERENCENUMBER=0
\def\REF#1{\expandafter\ifx\csname RF#1\endcsname\relax
               \global\advance\REFERENCENUMBER by 1
               \expandafter\xdef\csname RF#1\endcsname
                   {\the\REFERENCENUMBER}\fi}
\def\reftag#1{\expandafter\ifx\csname RF#1\endcsname\relax
               \global\advance\REFERENCENUMBER by 1
               \expandafter\xdef\csname RF#1\endcsname
                      {\the\REFERENCENUMBER}\fi
             \csname RF#1\endcsname\relax}
\def\ref#1{\expandafter\ifx\csname RF#1\endcsname\relax
               \global\advance\REFERENCENUMBER by 1
               \expandafter\xdef\csname RF#1\endcsname
                      {\the\REFERENCENUMBER}\fi
             [\csname RF#1\endcsname]\relax}
\def\refto#1#2{\expandafter\ifx\csname RF#1\endcsname\relax
               \global\advance\REFERENCENUMBER by 1
               \expandafter\xdef\csname RF#1\endcsname
                      {\the\REFERENCENUMBER}\fi
           \expandafter\ifx\csname RF#2\endcsname\relax
               \global\advance\REFERENCENUMBER by 1
               \expandafter\xdef\csname RF#2\endcsname
                      {\the\REFERENCENUMBER}\fi
             [\csname RF#1\endcsname--\csname RF#2\endcsname]\relax}
\def\refs#1#2{\expandafter\ifx\csname RF#1\endcsname\relax
               \global\advance\REFERENCENUMBER by 1
               \expandafter\xdef\csname RF#1\endcsname
                      {\the\REFERENCENUMBER}\fi
           \expandafter\ifx\csname RF#2\endcsname\relax
               \global\advance\REFERENCENUMBER by 1
               \expandafter\xdef\csname RF#2\endcsname
                      {\the\REFERENCENUMBER}\fi
            [\csname RF#1\endcsname,\csname RF#2\endcsname]\relax}
\def\refss#1#2#3{\expandafter\ifx\csname RF#1\endcsname\relax
               \global\advance\REFERENCENUMBER by 1
               \expandafter\xdef\csname RF#1\endcsname
                      {\the\REFERENCENUMBER}\fi
           \expandafter\ifx\csname RF#2\endcsname\relax
               \global\advance\REFERENCENUMBER by 1
               \expandafter\xdef\csname RF#2\endcsname
                      {\the\REFERENCENUMBER}\fi
           \expandafter\ifx\csname RF#3\endcsname\relax
               \global\advance\REFERENCENUMBER by 1
               \expandafter\xdef\csname RF#3\endcsname
                      {\the\REFERENCENUMBER}\fi
[\csname RF#1\endcsname,\csname RF#2\endcsname,\csname
RF#3\endcsname]\relax}
\def\refand#1#2{\expandafter\ifx\csname RF#1\endcsname\relax
               \global\advance\REFERENCENUMBER by 1
               \expandafter\xdef\csname RF#1\endcsname
                      {\the\REFERENCENUMBER}\fi
           \expandafter\ifx\csname RF#2\endcsname\relax
               \global\advance\REFERENCENUMBER by 1
               \expandafter\xdef\csname RF#2\endcsname
                      {\the\REFERENCENUMBER}\fi
            [\csname RF#1\endcsname,\csname RF#2\endcsname]\relax}
\def\Ref#1{\expandafter\ifx\csname RF#1\endcsname\relax
               \global\advance\REFERENCENUMBER by 1
               \expandafter\xdef\csname RF#1\endcsname
                      {\the\REFERENCENUMBER}\fi
             [\csname RF#1\endcsname]\relax}
\def\Refto#1#2{\expandafter\ifx\csname RF#1\endcsname\relax
               \global\advance\REFERENCENUMBER by 1
               \expandafter\xdef\csname RF#1\endcsname
                      {\the\REFERENCENUMBER}\fi
           \expandafter\ifx\csname RF#2\endcsname\relax
               \global\advance\REFERENCENUMBER by 1
               \expandafter\xdef\csname RF#2\endcsname
                      {\the\REFERENCENUMBER}\fi
            [\csname RF#1\endcsname--\csname RF#2\endcsname]\relax}
\def\Refand#1#2{\expandafter\ifx\csname RF#1\endcsname\relax
               \global\advance\REFERENCENUMBER by 1
               \expandafter\xdef\csname RF#1\endcsname
                      {\the\REFERENCENUMBER}\fi
           \expandafter\ifx\csname RF#2\endcsname\relax
               \global\advance\REFERENCENUMBER by 1
               \expandafter\xdef\csname RF#2\endcsname
                      {\the\REFERENCENUMBER}\fi
        [\csname RF#1\endcsname,\csname RF#2\endcsname]\relax}
\def\refadd#1{\expandafter\ifx\csname RF#1\endcsname\relax
               \global\advance\REFERENCENUMBER by 1
               \expandafter\xdef\csname RF#1\endcsname
                      {\the\REFERENCENUMBER}\fi \relax}

%

\def\NP{{ Nucl.\ Phys.\ }}
\def\PL{{ Phys.\ Lett.\ }}
\def\PR{{ Phys.\ Rev.\ }}

\def\PRL{{ Phys.\ Rev.\ Lett.\ }}

\def\ZP{{ Z.\ Phys.\ }}

\magnification=1200
\hsize=16.0truecm
\vsize=24.0truecm
\baselineskip=13pt
\pageno=0
\def\la{\Lambda_{\rm QCD}}
\def\s_8{\sigma_8}
\def\Q{Q{\bar Q}}
{}~~~ \hfill CERN-TH/95-214  \par
{}~~~ \hfill BI-TP 95/30
\vskip 2.5truecm
\centerline{\bf CHARMONIUM COMPOSITION AND NUCLEAR SUPPRESSION}
\vskip 1.5 truecm
\centerline{\bf D.\ Kharzeev and H.\ Satz}
\bigskip\medskip
\centerline{Theory Division, CERN, CH-1211 Geneva, Switzerland}
\centerline{and}
\centerline{Fakult\"at f\"ur Physik, Universit\"at Bielefeld,
D-33501 Bielefeld, Germany}
\vskip 2 truecm
\centerline{\bf Abstract:}
\medskip
We study charmonium production in hadron-nucleus collisions through
the intermediate next-to-leading Fock space component $|(\c)_8g\rangle$,
formed by a colour octet $\c$ pair and a gluon. We estimate
the size of this state and show that its interaction with nucleons
accounts for the observed charmonium suppression in nuclear interactions.
\vfill
\noindent \hrule
{}~~~\par
\noindent CERN-TH/95-214
\par
\noindent BI-TP 95/30
\par
\noindent August 1995
\eject

\refadd{Bodwin}
Charmonium production in hadronic collisions inherently involves
different energy and time scales; for an extensive recent treatment,
see \ref{Bodwin}. The first step is the creation of a heavy $\c$ pair,
e.g., by gluon fusion; this takes place in a very short
time, $\t_{\rm pert} \simeq m_c^{-1}$. The pair is in a colour octet
state; to neutralise its colour and yield a resonance state of
\J~quantum numbers, it has to absorb or emit an additional gluon
(Fig.\ 1). The time $\t_8$ associated to this process is determined by
the virtuality of the intermediate $\c$ state. In the rest frame of
the $\c$, it is approximately \ref{KS1}
$$
\t_8 \simeq  {1\over \sqrt \Delta}, \eqno(1)
$$
where $\Delta \equiv [(p+k)^2-m_c^2]= 2pk$. For massless quarks,
this gives the familiar $1/k_T$ for the time uncertainty associated
with gluon emission/absorption; for charm quarks of (large)
mass $m_c$, we get
$$
\t_8 \simeq {1 \over \sqrt{2m_c k_0}} \eqno(2)
$$
where $k_0$ is the energy of the additional gluon. If it is
sufficiently soft, $\t_8 > \t_{\rm pert}$. For \J~production
at mid-rapidity of a nucleon-nucleon collision, the colour
neutralisation time becomes
$$
t_8 \simeq \t_8 [1+(P_A/2m_c)^2]^{1/2} \eqno(3)
$$
in the rest frame of target or projectile nucleon, with $P_A$
denoting the momentum of the $\c$ pair in this frame. As seen from the
nucleon, colour neutralisation of fast $\c$ pairs will thus take a
long time. Equivalently, a fast $\c$ travels in the time $t_8$ a long
distance,
$$
d_8 \simeq \t_8 (P_a/2m_c), \eqno(4)
$$
in the rest frame of the nucleon. On the basis of the process shown
in Fig.\ 1, this seems to imply the existence of a coloured $\c$
state of well-defined quantum numbers
over times or distances much greater than the confinement
scale of about 1 fm (corresponding to $\la^{-1}$ with $\la\simeq
0.2$ GeV). This problem is particularly evident for \J~production
at low transverse momentum, for which the additional gluon has to
be quite soft; in this case, however, perturbative arguments become
in any case questionable. The problem seems avoidable for production
at sufficiently high $p_T$, i.e., for large enough $k_0$. Hence
the colour singlet model \refs{CS}{Schuler},
which treats the complete colour
singlet formation process in Fig.\ 1 within perturbative QCD, was
usually restricted to high $p_T$ production.
The low $p_T$ problem was until now usually ``solved" by
noting that inclusive \J~production occurs in the colour field
of the collision, leaving the form of the colour neutralisation
unspecified (``colour evaporation" \refs{CE}{Quarko}).
\par
Recent data \ref{Fermi} have shown, however, that also at high
$p_T$ non-perturbative long-time features seem to be essential
for charmonium production \refs{Bodwin}{Braaten}. The most important
outcome of these studies (see \ref{Mangano} for a recent review)
is that higher Fock space components of charmonium states play a
dominant role in their production. We thus decompose the \J~state
$|\psi \rangle$
$$
|\psi\rangle = a_0 |(\c)_1\rangle + a_1 |(\c)_8 g\rangle
+ a_2 |(\c)_1 gg \rangle + a'_2|(\c)_8 gg \rangle + ... \eqno(5)
$$
into a pure $\c$ colour singlet component ($^3S_1$), into a component
consisting of $\c$ colour octet ($^1S_0$ or $^3P_J$) plus a gluon,
and so on. The higher
Fock space coefficients correspond to an expansion in the relative
velocity $v$ of the charm quarks. For the wave function of the \J, the
higher components thus correspond to (generally small) relativistic
corrections. For \J~production, however, their role can become
decisive. While in short-time production the lowest component is
the most important, in long-time processes the next higher term becomes
dominant.
\par
Analogous decompositions hold for the other charmonium states
\refs{Bodwin}{Braaten}. In all
cases, the first higher Fock space state consists of a colour octet
$\c$ plus a gluon. For the \P, the next-to-leading terms again contain
a colour octet ($^3P_J$ or $^1S_0$ $\c$)
plus a gluon; for the $\chi$'s,
a $^3S_1$ colour octet $\c$ is combined with a gluon.
\par
This sheds some light onto the unspecified colour evaporation process.
When the colour octet $\c$ leaves the field of the nucleon in which
it was produced, it will in general neutralise its colour by combining
non-perturbatively with an additional collinear gluon, thus producing
the $(\c)_8 g$ component of the \J~or the other charmonium
states (Fig.\ 2). A necessary prerequisite for this is the small
size of the $\c$, due to the heavy quark mass; it is only because of
this that the soft gluon interacts with the $\c$ as a colour octet
and not with the individual quarks.
After the ``relaxation time" $\t_8$, the $(\c)_8 g$
will then absorb the accompanying gluon to revert to the dominant
$(\c)_1$ charmonium mode (Fig.\ 3).
Note that we are here considering those $\c$
pairs which will later on form charmonia. The $(\c)_8$ could as well
neutralise its colour by combining with a light quark-antiquark pair,
but this would result predominantly in open charm production.
\par
The production of charmonia in a kinematic regime involving long
time scales thus implies the production of the composite and hence
extensive state $(\c)_8g$. Its intrinsic transverse size
$r_8$ can be estimated through the uncertainty in the transverse
momentum induced in the charm quark when it absorbs the accompanying
gluon to go from $|(\c)_8g \rangle$
to the basic Fock component $|(\c)_1\rangle$
(Fig.\ 3). From the non-relativistic form for heavy quarks,
$$
{p^2 \over 2m_c} \simeq k_0, \eqno(6)
$$
where $p$ is the quark transverse momentum in the $(\c)_8 g$ cms,
we obtain from the lowest allowed gluon energy $k_0=\la$
the intrinsic size
$$
r_8 \simeq {1\over \sqrt{2m_c\la}} \simeq 0.25~{\rm fm}. \eqno(7)
$$
Since this size is determined only by the $(\c)_8 g$ composition of the
next-to-leading Fock space state and the gluon momentum cut-off in
confined systems, it is the same for all charmonium states.
In general, the time and momentum uncertainty would be given
by the binding energy and the size of the state, i.e., for the \J~by
the mass
difference $\e_0 =2M_D - M_{\psi} \simeq 0.64$ GeV between it and open
charm. This would imply different sizes for the $(\c)_8 g$
component of different resonance states; for a related discussion,
see \ref{Mueller}. In the case of all charmonium and
bottonium states, however, the common confinement cut-off $\sqrt{2m_Q
\la}$ is more stringent and hence the relevant one.
\par
\refadd{Weise}
\refadd{Gavin}
\refadd{Heinz}
\refadd{Bores}
\refadd{Dolej}
We now want to study the effect of these considerations on \J~production
in proton-nucleus collisions. For low $P_T$ production, with $k_0 \simeq
\la$ in Eq.\ (2), $\t_8$ would exceed the size of the even heavy nuclei
for \J's of sufficiently high lab momenta. Maintaining an approach based
on the colour singlet model would thus require a colour octet
$\c$ to pass
through the entire nuclear medium \ref{KS1}\refto{Weise}{Bores}.
The composition of the
colour-neutralising cloud needed for this was so far undetermined, and
hence estimates for the resulting cross sections were generally obtained
by assuming such a dressed $(\c)_8$ to be of hadronic size
\refss{KS1}{Weise}{Dolej}.
It was also left open why interactions with
the surrounding cloud would not destroy the spatial and quantum number
structure of the $(\c)_8$. Quarkonium production through
higher Fock space components now provides a specific description of the
$(\c)_8$ passage through the nucleus. The system leaves the nucleon
in whose field it was formed as a colour singlet $(\c)_8 g$
and continues as such through the nuclear medium. The size of this
charmonium state $(\c g) \equiv |(\c)_8g \rangle$  is given
by Eq.\ (7) and is thus essentially that of a \J. Also its interaction
with hadrons is similar to the \J-hadron interaction \ref{KS3}, but with
two important distinctions. The gluon exchanged between the two
colliding systems will now couple predominantly to the gluon
or to the $(\c)_8$ component of the $(\c g)$. Since both
are colour octets, in contrast to the colour triplet components of the
\J, the coupling is correspondingly enhanced
by a factor 9/4. Such an interaction will render the $(\c)_8 g$ system
coloured. Due to the repulsive one-gluon exchange interaction, the
colour octet $(\c_8)$ is not bound, in contrast to the
colour singlet $(\c)_1$. Moreover,
the probability of the $(\c)_8$ encountering another
collinear gluon to again form a colour singlet system of
\J~quantum numbers is minimal. Hence any $(\c g)$ interaction will
generally lead to its break-up, so that there is no threshold
factor. The cross section for $(\c g)$-hadron interactions is thus
just the geometric \J-hadron cross section, $\sigma_{\psi N}$,
increased by the enhanced coupling,
$$
\sigma_{(\c g) N} = {9\over 4} \sigma_{\psi N}.\eqno(8)
$$
Short-distance QCD calculations \ref{KS3} lead to
$\sigma_{\psi N} \simeq$ 2.5 - 3  mb, in good agreement
with photoproduction data \ref{photo}. From this we get
$\sigma_{(\c g) N} \simeq$ 6 - 7 mb. Because of the general nature of
the arguments leading to this value, it holds equally for the
interactions of the next-to-leading Fock space components of \P~and
$\chi$ with hadrons.
\par
This cross section, while larger than the high energy \J-hadron
cross section by about a factor two, is very much smaller than the
hadronic value of 20 - 50 mb previously assumed for the colour octet
$\c$ passing through the nucleus \refss{KS1}{Weise}{Dolej}.
This has immediate consequences. The mean free path of the
$(\c g)$ in nuclear matter, $\lambda_{(\c g)}=1/n_0
\sigma_{(\c g) N} \simeq 10$ fm is larger than the radius of
even the heaviest nuclei. Moreover, since the $\c$ combines with
an already existing collinear gluon, there is no coherence length
associated with any $(\c g)$ formation. Hence shadowing \refs{GS1}{KS2}
is excluded as dominant quarkonium suppression mechanism in $p-A$
collisions.
\par
In passing through the nucleus, the small physical state $(\c g)$
will interact incoherently with the nucleons along its trajectory. The
charmonium production probability on nuclei relative to that on
nucleons, valid for \J, $\chi$ and \P~production, thus becomes
$$
S_A = exp\{-n_0\sigma_{(\c g) N} L\} , \eqno(9)
$$
where $L$ denotes the length of the path through nuclear matter
of standard density $n_0=0.17$ fm$^{-3}$, and
$\sigma_{(\c g) N} \simeq$ 6 - 7 mb is the inelastic
$(\c g)$-nucleon cross section obtained above.
\par
With Eq.\ (9) we have derived the Gerschel-H\"ufner fit \ref{Gerschel},
introduced as a phenomenological description of hadron-nucleus data
on \J~production. In \ref{Gerschel}, it was attempted to interpret
the cross section entering in Eq.\ (9) as the physical \J-hadron
cross section, which led to theoretical as well as  experimental
problems. The fit value was 5 - 7 mb; both short distance QCD \ref{KS3}
and photoproduction experiments \ref{photo} give a \J-hadron cross
section smaller by at least a factor two. Moreover, $p-A$ data
\refs{E772}{Carlos} lead to
exactly the same suppression of \P~and \J~production for all $A$,
with an $A$-independent production ratio \P/(\J)$\simeq 0.15$.
Since the geometric size of the ground state \J~and the next
radial excitation differ by more than a factor four, an equal
suppression contradicts the interaction of fully
formed physical resonances. We find here that all aspects of
the observed suppression arise naturally in charmonium production
through the next higher Fock space component $(\c g)$. A composite
state $(\c g)$, which is of the same size for all charmonia,
passes through the nuclear medium and hence leads
to equal \P~and \J~suppression. The value of the cross section
$\sigma_{(\c g)N}$ thus obtained is in good agreement
with that obtained from a fit to the data \ref{Gerschel}.
\par
\par
The same argumentation also provides the suppression of bottonium
production in $p-A$ collisions. The radius of the $(\b)_8g$ state
is (see Eq.\ (7)) a factor $\sqrt{m_c/m_b} \simeq
\sqrt 3$ smaller than that of the $(\c g)$, so that
$\sigma_{(\b g) N} \simeq (1/3)\sigma_{(\c g)N} \simeq 2$ mb.
\par
To illustrate how well both \J~and \U~production in $p-A$ collisions
are described by this scenario, we show in Fig.\ 4 recent high energy
data at $\sqrt s \simeq 20$ \ref{Carlos}, 30 \ref{Fredj} and 40
GeV \ref{E772}. With an average path length
$L_A=(3/4)[(A-1)/A][1.12~A^{1/3}]$ we obtain excellent agreement for
$\sigma_{(\c g)N} \simeq 6$ mb and $\sigma_{(\b g) N} \simeq 2$ mb.
In \ref{Gerschel} it was shown that \J~production data from $\pi-A$
collisons lead to very similar values for the cross section in Eq.\
(9).
\par
With quarkonium suppression in hadron-nucleus collisions thus
accounted for in terms of interaction between $(\Q)_8g$ states
and nucleons, we can now also consider nucleus-nucleus interactions.
For collision energies of $\sqrt s \simeq 20$ GeV, the centers of
the two colliding nuclei have at time $t_8$ separated
in the cms by about 5 fm.
With the nuclei Lorentz-contracted to a thickness of
about 0.5 fm or less, this means that a \J~produced at mid-rapidity
in the cms has experienced in its early phase an effect
corresponding to that obtained by superimposing the passages through
the two nuclei \ref{Gerschel}. Hence the charmonium suppression
now is given by
$$
S_{(A-B)} = exp\{-n_0\sigma_{(\c g)N}(L_A+L_B)\} , \eqno(10)
$$
again in accord with the phenomenological fit of \ref{Gerschel}.
The path length $L=L_A+L_B$
in Eq.\ (10) varies with impact parameter and hence
with the associated transverse energy $E_T$ produced in the collision.
A relation between $L$ and $E_T$ can thus be obtained through a
detailed study of the collision geometry \ref{Salmeron}.
An alternative is
given by determining $L$ through the broadening of the average
transverse momentum of \J~or Drell-Yan dileptons, since
this broadening also depends on the average path length \ref{NA38G}.
In Fig.\ 5 we show the resulting values \ref{Carlos}, normalised to the
$(\c g)$ suppression of Eqs. (8) and (9), for both $p-A$
and $S-U$ data. We conclude that also the \J~suppression observed
so far in $S-U$ collisions is completely accounted for by $(\c
g)$ suppression in standard nuclear matter.
\par
The comparison of \J~and \P~production in nucleus-nucleus collisions
provides a test for the presence of a medium in the later stages.
As long as the interaction leading to charmonium suppression is
determined by the $(\c g)$ state, \J~and \P~must be equally
suppressed. To make \J~and \P~suppression different,
the medium must see the fully formed resonances and distinguish
between them. In Fig.\ 5 we have included also the \P~suppression
divided by the $(\c g)$ suppression (8) \refs{NA38n}{Carlos}.
It is evident that in $S-U$ collisions \J~and \P~are not affected
equally, the \P~being much stronger suppressed. This establishes
the presence of a medium at a time late enough for fully formed
charmonium resonances to exist. On the other hand, this medium
breaks up only the \P, leaving the \J~unaffected. It was shown
\ref{KS3} that interactions with hadrons in the range of present
collision energies cannot dissociate a \J, while for the much
more loosely bound \P~this is readily possible. We therefore
conclude that the medium probed by charmonium production in present
$S-U$ collisions is of hadronic nature, i.e., confined; it could
consist either of stopped nucleons or of secondary hadrons
produced in the collision \ref{Wong}. Note that we include
the \P~data as function of $L$ in Fig.\ 5 simply to show the
additional suppression present in this case. It is not at all
clear that $L$ is a meaningful variable for the effect of such a
confined environment on charmonia. We also note that the
appearent absence of any effect of this medium on the observed \J,
even though this is to about 40 \% produced through $\chi$ decay,
is in accord with the short distance QCD form of the $\chi$-hadron
cross section \ref{KS-CD}.
\par
To establish colour
deconfinement, either in equilibrium or pre-equilibrium
systems, nucleus-nucleus collisions have to produce a \J~suppression
beyond that given by Eq.\ (10), i.e., beyond that found in $p-A$
collisions \ref{Gerschel}, and different from \P~suppression.
If such an additional suppression
were found, the results for inelastic \J-hadron collisions
obtained from short distance QCD \ref{KS1} would rule out a confined
medium, and the difference in \J~and \P~suppression would exclude
$(\c g)$ interactions as a source.
\par
In summary: we study quarkonium production in hadron-nucleus
collisions through the intermediate next-to-leading Fock space state
consisting of a colour octet $\c$ and a gluon. We estimate the
inelastic $(\c g)$-nucleon cross section and use this to
\par
\item{--}{exclude shadowing as main origin for the quarkonium suppression
observed in hadron-nucleus interactions;}
\item{--}{derive the Gerschel-H\"ufner fit describing such suppression,
both for $\c$ and $\b$ states;}
\item{--}{conclude that the \J~suppression observed in $S-U$
collisions is fully accounted for by early $(\c g)$ interactions with
standard nuclear matter;}
\item{--}{conclude that the stronger \P~suppression found in $S-U$
collisions is due to an additional confined medium present at a later
stage.}
\par\noindent
It will be interesting to see if the results of forthcoming $Pb-Pb$
collisions can provide first indications for deconfinement -- in
particular, if they show
a stronger \J~dissociation than accounted for by $(\c g)$ interactions.
\bigskip
\centerline{\bf Acknowledgements}
\medskip
We thank A. Capella, C. Gerschel, A. Kaidalov, C.
Lourenco and M. Mangano for stimulating and helpful discussions.
The support of the German Research Ministry BMFT under contract
06 BI 721 is gratefully acknowledged.
\bigskip
\medskip
\centerline{\bf References}
\medskip
\item{\reftag{Bodwin})}{G. T. Bodwin, E. Braaten and G. P. Lepage,
\PR D 51 (1995) 1125}
\par
\item{\reftag{KS1})}{D. Kharzeev and H. Satz, \ZP C 60 (1993) 389.}
\par
\item{\reftag{CS})}{C. H. Chang, \NP B 172 (1980) 425;\hfill\break
E. L. Berger and D. Jones, \PR D 23 (1981) 1521;\hfill\break
R. Baier and R. R\"uckl, \PL B 102 (1981) 364 and \ZP C 19 (1983) 251.}
\par
\item{\reftag{Schuler})}{For a recent survey, see G. A. Schuler,
``Quarkonium Production and Decays", CERN-TH.7174/94, 1994, to
appear in Physics Reports.}
\par
\item{\reftag{CE})}{M. B. Einhorn and S. D. Ellis, \PR D12 (1975)
2007;\hfill\break
H. Fritzsch, \PL 67B (1977) 217;\hfill\break
M. Gl\"uck, J. F. Owens and E. Reya, \PR D17 (1978) 2324;\hfill\break
J. Babcock, D. Sivers and S. Wolfram, \PR D18 (1978) 162.}
\par
\item{\reftag{Quarko})}{For a recent survey, see
R. V. Gavai et al., ``Quarkonium Production in Hadronic Collisions",
CERN-TH.7526/94, 1994, to appear in Int. J. Mod. Phys. A.}
\par
\item{\reftag{Fermi})}{See e.g. V. Papadimitriou (CDF), ``Production
of Heavy Quark States at CDF", Preprint Fermilab-Conf-95-128 E,
March 1995; \hfill \break
L. Markosky (D0), ``Measurements of Heavy Quark Production at D0",
Preprint Fermilab-Conf-95-137 E, May 1995.}
\par
\item{\reftag{Braaten})}{E. Braaten and S. Fleming, \PRL 74 (1995)
3327.}
\par
\item{\reftag{Mangano})}{M. L. Mangano, ``Phenomenology of Quarkonium
Production in Hadronic Collisions", CERN-TH/95-190.}
\par
\item{\reftag{Mueller})}{A. H. Mueller, \NP B 415 (1994) 373.}
\par
\item{\reftag{Weise})}{G. Piller, J. Mutzbauer and W. Weise, \ZP A 343
(1992) 247, and \NP A 560 (1993) 437.}
\par
\item{\reftag{Gavin})}{S. Gavin and J. Milana, \PRL 68 (1992) 1834.}
\par
\item{\reftag{Heinz})}{R. Wittmann and U. Heinz, \ZP C 59 (1993) 77.}
\par
\item{\reftag{Bores})}{K. Boreskov et al., \PR D47 (1993) 919.}
\par
\item{\reftag{Dolej})}{J. Dolej\v si and J. H\"ufner, \ZP C 54 (1992)
489.}
\par
\item{\reftag{KS3})}{D. Kharzeev and H. Satz, \PL B 334 (1994) 155.}
\par
\item{\reftag{photo})}{S. D. Holmes, W. Lee and J. E. Wiss, Ann. Rev.
Nucl. Part. Sci. 35 (1985) 397.}
\par
\item{\reftag{GS1})}{S. Gupta and H. Satz, \ZP C 55 (1992) 391.}
\par
\item{\reftag{KS2})}{D. Kharzeev and H. Satz, \PL B 327 (1994) 361.}
\par
\item{\reftag{Gerschel})}{C. Gerschel and J. H\"ufner,
\ZP C 56 (1992) 171.}
\par
\item{\reftag{E772})}{D. M. Alde et al., Phys. Rev. Lett. 66 (1991)
133;\hfill\break
D. M. Alde et al., Phys. Rev. Lett. 66 (1991) 2285.}
\par
\item{\reftag{Carlos})}{C. Lourenco (NA38/51),
``Recent Results on Dimuon Production from the NA38 Experiment",
CERN-PPE/95-72, May 1995, and Doctorate Thesis,
Universidade T\'ecnica de Lisboa, Portugal, January 1995.}
\par
\item{\reftag{Fredj})}{L. Fredj (NA38/51), Doctorate Thesis,
Universit\'e de Clermont-Ferrand, France, September 1991.}
\par
\item{\reftag{Salmeron})}{R. Salmeron, \NP B 389 (1993) 301.}
\par
\item{\reftag{NA38G})}{C. Baglin et al., \PL B 268 (1991) 453.}
\par
\item{\reftag{NA38n})}{C. Baglin et al., \PL B 345 (1995) 617.}
\par
\item{\reftag{Wong})}{C.-Y. Wong, ``Suppression of \P~and \J~in
High Energy Heavy Ion Collisions", Oak Ridge Preprint ORNL-CTP-95-04,
June 1995.}
\item{\reftag{KS-CD})}{D. Kharzeev and H. Satz, ``Colour Deconfinement
and Quarkonium Dissociation", Preprint CERN-TH/95-117, May 1995; to
appear in R. C. Hwa (Ed.), {\sl Quark-Gluon Plasma II}, World
Scientific Publ. Co., Singapore.}
\bigskip
\medskip
\centerline{\bf Figure Captions}
\medskip
\leftskip=0.6 truecm
\item{Fig.\ 1:}{\J~production through gluon fusion.}
\bigskip
\item{Fig.\ 2:}{\J~production via $(\c g)$ colour singlet.}
\bigskip
\item{Fig.\ 3:}{Transition from $(\c g)$ colour singlet to $\c$ colour
singlet.}
\bigskip
\item{Fig.\ 4:}{\J~and \U~suppression in $p-A$ collisions; data
\refto{E772}{Fredj} are compared to $(\c g)$ suppression (Eq.\ (9))
in nuclear matter and to the corresponding form for $(\b g)$.}
\bigskip
\item{Fig.\ 5:}{\J, \P~and \U~suppression in $p-A$ and $S-U$
collisions, normalised to $(\c g)$ and $(\b g)$ suppression in
nuclear matter.}
\par
\vfill\bye